\documentclass[epj]{svjour}
\usepackage{graphics}
\begin{document}
\title{Surface effects in nanoparticles: application to maghemite $\gamma $-Fe$_{2}$O$_{3}$}
\author{H. Kachkachi\inst{1}\thanks{\emph{Corresponding author}
kachkach@physique.uvsq.fr}, A.\ Ezzir\inst{1}, M.\ Nogu\`{e}s\inst{1}, \and E. Tronc\inst{2}
}                     
%
%
\institute{
LMOV-CNRS UMR 8634, 
Universit\'{e} de Versailles St. Quentin,
45, avenue des Etats-Unis,
78035 Versailles Cedex, France 
\and 
LCMC-CNRS URA\ 1466, 
Universit\'{e} Pierre \& Marie Curie, 
4 place Jussieu, 75252 Paris Cedex, France}
\date{Received: date / Revised version: date}
%
\abstract{
We present a microscopic model for nanoparticles, of the maghemite ($\gamma $%
-Fe$_{2}$O$_{3}$) type, and perform classical Monte Carlo
simulations of their magnetic properties. On account of M\"{o}ssbauer
spectroscopy and high-field magnetisation results, we consider a particle as
composed of a core and a surface shell of constant thickness.
The magnetic state in the particle is described by the anisotropic classical
Dirac-Heisenberg model including exchange and dipolar interactions and bulk
and surface anisotropy. We consider the case of ellipsoidal (or spherical)
particles with free boundaries at the surface. Using a surface
shell of constant thickness ($\sim 0.35$ nm) we vary the particle size and
study the effect of surface magnetic disorder on the thermal and spatial behaviors
of the net magnetisation of the particle. We study the shift in the surface
``critical region'' for different surface-to-core ratios of the exchange
coupling constants. It is also shown that the profile of the local
magnetisation exhibits strong temperature dependence, and that surface anisotropy
is reponsible for the non saturation of the magnetisation at low temperatures.
} 
\maketitle
\section{Introduction}
\label{intro}
Surface effects in a nanoparticle are of great importance since they
dominate the magnetic properties and become more important with decreasing
size of the particle. In particular, the picture of a single-domain magnetic
particle where all spins are pointing into the same direction, leading to
coherent relaxation processes, is no longer valid when one considers the
effect of misaligned spins on the surface on the global magnetic properties
of the particle. Indeed, even for strong exchange interactions the surface
is responsible for coordination defects, and in a particle of radius of the
order of $4$ nm, $50\%$ (in the $\gamma $-Fe$_{2}$O$_{3}$ nanoparticles) of
atoms lie on the surface, and therefore the effect of the latter cannot be
neglected.

The magnetisation near the surface is generally lower than in the interior.
This effect leads to a thermodynamic perturbation in exchange interactions
near the surface, which can be sizable at high temperatures. In a finite
system, the effect of temperature introduces, in addition to the usual
renormalization of the spin-wave spectrum, a spatial dependence of the
magnetisation (see \cite{Wolfram et al.} and references therein for the case
of semi-infinite cubic crystals). Indeed, the symmetry breaking at the
surface results in a surface anisotropy. In most cases, this happens to be
strong enough as to compensate for the work needed against the exchange
energy that prefers full alignment, and it is conceivable then that the
magnetisation vector will point along the bulk easy axis in the core of the
particle, and will then gradually turn into a different direction when it
approaches the surface. In addition to the exchange interactions, which are
the strongest interactions between atoms in a magnetic system, there are
also the purely magnetic dipole interactions between the magnetic moments of
the atoms and the interactions between the magnetic moments and the electric
field of the crystal lattice (spin-orbit interactions). The last two types
of interactions are relativistic in origin and therefore correspond to
energies that are much smaller than the exchange energy on a short length
scale, but as they are long range interactions they lead, in general, to non
negligible contributions. Indeed, these interactions play an important role
in that they introduce a preferred direction in the system as they
correspond to a Hamiltonian (see below) that is not invariant under rotation
operations. In other words, they lead to the appearance of an anisotropy
energy, i.e. the dependence of the total energy of the system on the
direction of magnetisation. Moreover, the most important feature of the
relativistic interactions is that they change the magnetic moment of an atom
from one site to another, and hence introduce a non uniform spatial
distribution of the magnetic moment in the system. The treatment of the
dipole-dipole interactions \cite{Akhiezer} leads to two energy terms
corresponding to the volume and surface charges. In a small magnetic system,
such as a nanoparticle, only the second contribution is important since it
accounts for the shape anisotropy, and the former becomes negligible as one
integrates over a small volume (see a detailed discussion in \cite{Aharoni}
and references therein).

Therefore, it is necessary to take into account all different contributions
to the energy, exchange and dipolar interactions, bulk and surface
anisotropies, in any study of spatial distributions of the magnetisation in
a small magnetic system.

One of our ultimate goals is to understand the effect of surfaces on the thermodynamic
and spatial behaviors of the magnetisation in small systems, since this is
also of crucial importance to the study of the dynamics of such systems
where the surface effects are considerable. In this case, the dynamics is
rendered more complicated by the additional (metastable or stable) magnetic
states corresponding to the surface configurations.

\begin{figure*}
\begin{center}
\resizebox{0.75\textwidth}{!}{%
  \includegraphics{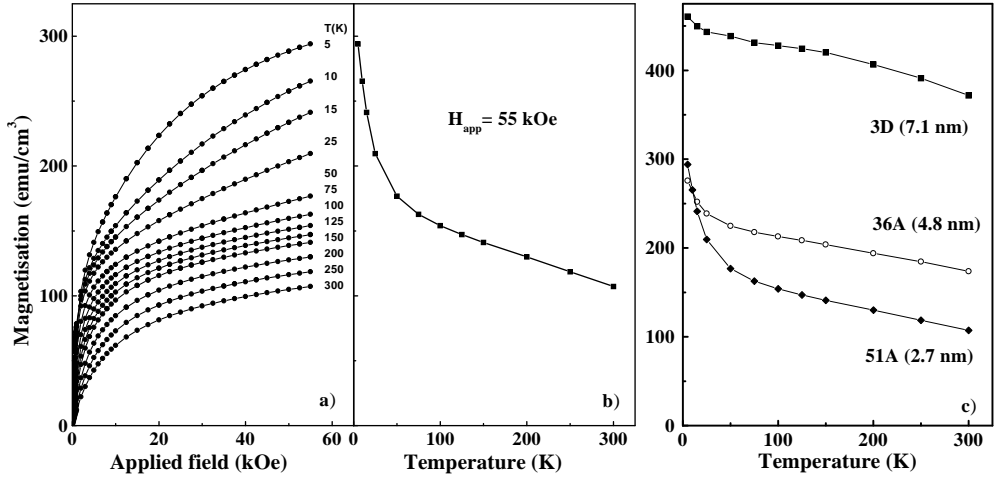}
}
\end{center}
\vspace*{-2.5cm}       
\caption{a) Magnetisation as a function of the magnetic field of a diluted assembly
of $\gamma $-Fe$_{2}$O$_{3}$ nanoparticles with a mean diameter of $2.7$ nm.
b) Thermal variation of the magnetisation extracted from a) at a field of $55$ kOe. 
c) Thermal variation of the magnetisation in a field of $55$ kOe for three
samples with different mean diameters ($2.7,4.8,7.1$ nm).}
\label{fig:1}       
\end{figure*}

According to M\"{o}ssbauer spectroscopy performed by Morrish et al. (see \cite{Haneda}
for a review , and also the seminal work by Coey \cite{Coey}), the
iron cations at the surface of the $\gamma $-Fe$_{2}$O$_{3}$ particles have
a noncollinear magnetic structure from 4.2 K to room temperature.
The M\"{o}ssbauer-spectroscopy analysis performed in \cite{Tronc
et al.} on $\gamma $-Fe$_{2}$O$_{3}$ particles show that the spectrum
contains two components, one associated with the bulk and the other with
iron atoms on the surface. The latter seems to disappear at temperatures in
the range $30-75$ K, depending on the mean diameter of the particle
assembly. From these M\"{o}ssbauer-effect analyses it was also inferred that
the surface shell has a thickness of about $0.35$ nm, independently of the
particle volume. This fact will be used in our simulations to determine the
ratio of the surface and core number of atoms.
In addition, measurements of the magnetisation at high fields performed on
the $\gamma $-Fe$_{2}$O$_{3}$ nanoparticles \cite{Ezzir} (see also \cite
{Chen et al.} for cobalt particles) have shown that the magnetisation is
strongly influenced by the surface effects, depending on the particle size.
In Figs.~\ref{fig:1} we show the field and thermal variation of the magnetisation for
different particle sizes. In Fig.~\ref{fig:1}a we see that there is a sudden increase
of the magnetisation as a function of the applied field when the temperature
reaches $70$ K, and that the magnetisation does not saturate at the highest
field value, i.e. 5.5 T. In Fig.~\ref{fig:1}b there is an important increase of the
magnetisation at low temperatures. In Fig.1c the thermal behaviour of the
magnetisation at 5.5 T is shown for samples with different mean diameter.
There we see that the smaller is the mean diameter of the particle the more
important is the increase of the magnetisation at very low temperatures.
This can be explained by the fact that the surface component corresponds to
a state with canted atomic moments at low temperatures \cite{Coey} \cite
{Tronc et al.}.
%
%
Furthermore, open hystereses were observed \cite{Kodama} in the nickel
ferrite particles (NiFe$_{2}$O$_{4}$) up to fields of $16$ T, while the
anisotropy field of the bulk sample is utmost $4\times 10^{-2}$ T. The
authors consider that the core spins are colinear whereas those on the
surface are misaligned. In the case of $\gamma $-Fe$_{2}O_{3}$ particles,
however, no such irreversibility in the magnetisation hysteresis loop was
observed \cite{Tronc et al.}, \cite{Kodama}.

Therefore, according to M\"{o}ssbauer spectroscopy and high-field
magnetisation measurements, we may think of a particle as containing a
magnetically disordered surface of a certain width and a core more or less ordered
depending on the size of the particle. The net magnetisation of the particle
is given by the (weighed) sum of the contributions from the surface and the core. The
increase of the surface contribution at low temperatures becomes more important
with decreasing size of the particle, as shown in Fig.1c.

In this work, we compute the thermal behavior of the different contributions
to the magnetisation due to the surface and core for different values of
exchange couplings and different surface-to-volume ratios of the number of spins. 
We also compute the spatial variation of the magnetisation at different temperatures 
and the specific heat for different sizes. All calculations are performed in zero
magnetic field.
%
%
\section{Model for magnetic particles}
\label{Model}
\subsection{Hamiltonian}
\label{Hamiltonian}
As was discussed in the introduction, it is well established \cite{Akhiezer}%
, \cite{Aharoni} that any theory dealing with the spatial magnetisation
distribution must consider all three energy contributions, exchange,
anisotropy and magnetostatic, and that there is no obvious reason for
neglecting any one of these. However, it should be emphasised that in very small
particles the exchange interactions are too strong to allow for subdivision
into domains, so that the magnetostatic energy term is considered here only
in order to account for the shape anisotropy in the case of ellipsoidal
particles. It should also be stressed, as discussed in the introduction,
that because of the surface anisotropy and the symmetry breaking at the
surface, there is no reason to consider the exchange interactions on the
surface as equal to those in the core, and thereby the particle may acquire
a non uniform distribution of the magnetisation.

Therefore, to take account of all these energy contributions, we propose a
model, based on the classical anisotropic Dirac-Heisenberg (exchange and dipolar)
Hamiltonian, describing the spinel structure of the (spherical or
ellipsoidal) ferrimagnetic nanoparticles (e.g., $\gamma $-Fe$_{2}$O$_{3}$).
In this article the calculations are performed using classical Monte Carlo simulations
of the magnetisation thermal behavior of the particle with open boundaries
at the surface.

So, the exchange, anisotropy and Zeeman contributions are given by the
following anisotropic Dirac-Heisenberg Hamiltonian 
\begin{eqnarray}
{\cal H}_{DH} =&-&\sum\limits_{i,{\bf n}}\sum\limits_{\alpha ,\beta
=A,B}J_{\alpha \beta }\;{\bf S}_{i}^{\alpha }\cdot {\bf S}_{i+{\bf n}%
}^{\beta }-K\sum\limits_{i=1}^{N_{t}}\left( {\bf S}_{i}\cdot {\bf e}%
_{i}\right)^{2} \nonumber \\
&-&(g\mu _{B})H\sum\limits_{i=1}^{N_{t}}{\bf S}_{i},  \label{1}
\end{eqnarray}
where $J_{\alpha \beta }$ (positive or negative) are the exchange coupling
constants between (the $\alpha ,\beta =A,B$) nearest neighbors spanned by
the unit vector ${\bf n}$; ${\bf S}_{i}^{\alpha }$ is the (classical) spin
vector of the $\alpha ^{th}$ atom at site $i;$ $H$ is the uniform field
applied to all spins (of number $N_{t}$) in the particle, $K>0$ is the
anisotropy constant and ${\bf e}_{i}$ the single-site anisotropy axis (see
definition below). A discussion of the core and surface anisotropy will be presented
below. In the sequel the magnetic field will be set to zero.

To the Dirac-Heisenberg Hamiltonian we add the pairwise long-range dipolar
interactions 
\begin{equation}
{\cal H}_{dip}=\frac{(g\mu _{B})^{2}}{2}\sum\limits_{i\neq j}\frac{\left( 
{\bf S}_{i}\cdot {\bf S}_{j}\right) R_{ij}^{2}-3\left( {\bf S}_{i}\cdot {\bf %
R}_{ij}\right) \cdot \left( {\bf R}_{ij}\cdot {\bf S}_{j}\right) }{R_{ij}^{5}%
}  \label{2}
\end{equation}
where $g$ is the Land\'{e} factor, $\mu _{B}$ the Bohr magneton and ${\bf R}%
_{ij}$ the vector connecting any two spins on sites $i$ and $j$ of the
particle, $R_{ij}\equiv \left\| {\bf R}_{ij}\right\| $.
\subsection{Method of simulation}
\label{Method}
The particle we consider here is a spinel with two different iron sites, a
tetrahedric Fe$^{3+}$ site (denoted by A) and an octahedric Fe$^{3+}$ site
(denoted by B). The nearest neighbor exchange interactions are (in units of
K) \cite{Smart}, \cite{Kodama}~: $J_{AB}/k_{B}\simeq -28.1$, $%
J_{BB}/k_{B}\simeq -8.6$, and $J_{AA}/k_{B}\simeq -21.0$. These coupling
constants are used in the Dirac-Heisenberg Hamiltonian ${\cal H}_{DH}$ in
order to model the phase transition from the paramagnetic to ferrimagnetic
order as the temperature is lowered down to zero through $T_{c}\simeq 906$
K. In the spinel structure an atom on site A has 12 nearest neighbors on the
sublattice B and 4 on the sublattice A, and an atom on site B has 6 nearest
neighbors on A and 6 on the B sublattice; the number of B sites is twice
that of sites A. The nominal value of the spin on sites A and B is $5/2$,
and this justifies the use of classical spins. 
We have also taken account of $\frac{1}{3}$ of lacuna for each two
B atoms randomly distributed in the particle.
The nanoparticle we have studied contains $N_{t}$ spins ($\simeq
10^{3}-10^{5}$), and its radius is in the range 2-3.5
nm. Our model is based on the hypothesis that the particle is composed of a
core of radius containing $N_{c}$ spins, and a surface
shell surrounding it that contains $N_{s}$ spins, so that $N_{t}=N_{c}+N_{s}$%
. Thus varying the size of the particle while maintaining the thickness of
the surface shell constant ($\sim 0.35$ nm), is equivalent to varying the
surface to total number of spins, $N_{st}=N_{s}/N_{t},$ and this allows us
to study the effect of surfaces of different contributions. All spins in the
core and on the surface are identical, but interact via the, {\it a priori},
different couplings depending on their locus in the whole volume. We will
consider both cases of identical interactions, and that of the general
situation with different interactions on the surface and in the core. Although we
treat only the crystallographically ``ideal'' surface, we do allow for
perturbations in the exchange constants on the surface. This is meant to
take into account, though in a somewhat phenomenological way, the possible
defects on the surface, and the possible interactions between the particles
and the matrix in which they are embedded. In \cite{Kodama} it was assumed
that the pairwise exchange interactions are of the same magnitude for the
core and surface atoms, but there was postulated the existence of a fraction
of missing bonds on the surface.
On the other hand, we consider that the exchange interactions
between the core and surface spins are the same as those inside the core. We
also stress that we are only concerned with non interacting particles, so we
ignore the effect of interparticle interactions on the exchange couplings at
the surface of the particle.

\begin{figure}
\resizebox{0.75\textwidth}{!}{%
\includegraphics{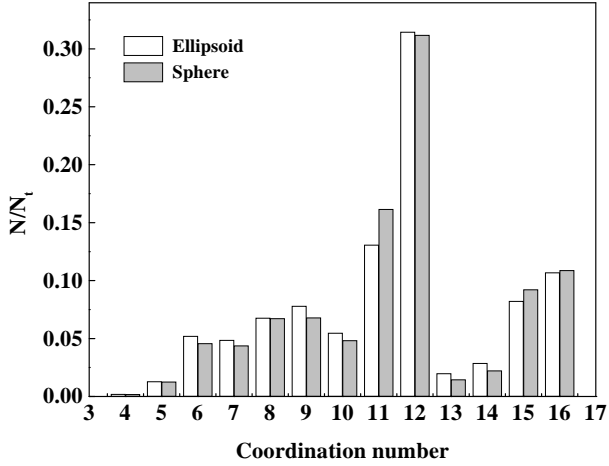}
}
\vspace{-2.5cm}      
\caption{Normalized distribution of the coordination number in a
spherical and ellipsoidal particle for $N_{st}=40\%$.}
\label{fig:2}       
\end{figure}

In our simulations we start with a regular box ($X\times Y\times Z$) of spins with the spinel
structure having the properties mentioned above, and then given the radius or total number of 
spins in the particle, we cut in a sphere or an
ellipsoid. We choose the center of the particle to lie at one of the lattice
sites. In all cases free boundary conditions are used, and thus a spin is considered as a 
core or surface spin depending on whether it has or not its full coordination number. 
Fig.~\ref{fig:2} shows the distribution of the coordination numbers in a
spherical and ellipsoidal particle.
%
%
Therefore, in our simulations each spin has a structure associated
with it that contains the coordinates of the corresponding site, the nature
of the site in the spinel structure (i.e. A or B), its locus (core or
surface), the number of its nearest neighbours, and the total number of such
sites. Thus, once the structure of each spin is defined, the spins outside
of the sphere or the ellipsoid are discarded. This accomplishes the  
simulation of the crystal structure and shape of the particle. Next, we proceed 
with the Monte Carlo calculations.

The classical Monte Carlo method based on the Metropolis algorithm is now a
standard method and detailed descriptions can be found in \cite{Binder & Krumbhaar}; 
so here we summarize only the main procedure. We calculate
the expectation value of a function $f(\left\{ {\bf S}_{i}\right\} )$ of the
spins ${\bf S}_{i}$%
\begin{equation}
\left\langle f\right\rangle =\frac{Tr\left[ \exp (-H/k_{B}T)f\right] }{%
Tr \exp (-H/k_{B}T) }  \label{3}
\end{equation}
by generating a Markov chain of spin configurations $\left\{ {\bf S}%
_{i}\right\} $ of the system and taking the average
\begin{equation}
\bar{f}=\frac{1}{L}\sum\limits_{c=1}^{L}f(\left\{ {\bf S}_{i}^{c}\right\} ).
\label{4}
\end{equation}
From any initial (random in our case) configuration $c$ we generate a trial
configuration by choosing the spin coordinates of a randomly chosen lattice
site $i$ by ($\alpha =x,y,z$)
\begin{equation}
S_{i}^{\alpha ,c+1}=\frac{S_{i}^{\alpha ,c}+X_{\alpha }.\Delta }{%
\sqrt{\sum\limits_{\alpha }\left(S_{i}^{\alpha ,c}+X_{\alpha
}.\Delta \right) ^{2}}},  \label{5}
\end{equation}
where $X_{\alpha }$ are random numbers satisfying $-1\leq X_{\alpha }\leq 1$. 
Then the energy change $\delta E$ produced by this move is calculated from
Eqs. (\ref{1}) and (\ref{2}). If $\delta E\geq 0,$ one compares $\exp
(-\delta E/k_{B}T)$ with a random number $\varsigma ,0\leq \varsigma \leq 1.$
If $\varsigma <\exp (-\delta E/k_{B}T)$ or $\delta E<0,$ the trial
configuration is accepted as new configuration, otherwise it is abandoned
and the old one is counted once more. It is shown \cite{Binder & Krumbhaar}
that this prescription leads to
\begin{equation}
\left\langle f\right\rangle =\lim_{L\rightarrow \infty }\bar{f}.  \label{6}
\end{equation}
Since one uses a finite $L,$ it is essential to choose the arbitrary step
parameter $\Delta$ suitably so as to exclude an appropriate number of initial
configurations from the average (\ref{4}), and to estimate the ``statistical
error'' in a reliable way. In our case, we always start with a configuration of 
randomly oriented spins and adjust $\Delta $ so
that 75\% of the moves were successful, and it turned out that about 3000
Monte-Carlo steps per spin were sufficient to yield an accuracy of a few
percent for the local magnetisation.

It is well known, of course, that spontaneous symmetry breaking, which is usually accompanied 
by a change from a disordered state at high temperatures to a spontaneously ordered state at
temperatures below the critical one, can occur only in the thermodynamic limit. 
In a finite magnetic system the magnetisation at zero field
\begin{equation}
{\bf m}(T,H=0) = \frac{1}{N}\sum\limits_{i=1}^{N}\left\langle {\bf S}_i\right\rangle_{T,H=0} 
\label{m}
\end{equation}
vanishes at all nonzero temperatures irrespective of the number $N$ of spins in the system. 
In particular, even above the critical temperature (of the bulk system) a small magnetisation is 
found by Monte Carlo calculations, due to fluctuations which in a finite system observed over a 
finite time have not completely averaged out \cite{Binder & Krumbhaar}. However, at 
very low temperatures there exist, even in an inhomogeneous system, clusters of spins aligned 
with respect to each other, and there should exist an {\it intrinsic magnetisation}. 
This is usually defined as follows
\begin{equation}
M=\sqrt{\left\langle \left(\frac{1}{N}
\sum \limits_{i=1}^{N}{\bf S}_{i}\right)^{2}\right\rangle}.  
\label{M}
\end{equation}
Of course, in a finite system the magnetisation $M$ is nonzero at all temperatures. 
Even at temperatures well above the critical temperature one still obtains $M\propto 1/\sqrt{N}$, 
and this leads to the appearance of a magnetisation tail at high temperatures. The 
magnetisation $M$ tends to that of the bulk system (infinite size) in the limit
$N \rightarrow \infty$. 
In the present case, we calculate the magnetisation defined in (\ref{M}) for the core and 
surface by summing over the corresponding spins.

{\bf Anisotropy}~{\bf energies~}: In both cases of a spherical and
ellispoidal particle, we consider a uniaxial anisotropy in the core and
single-site anisotropy on the surface. The easy axis in the core is chosen
along our $z$ reference axis, and the sites on the boundary have uniaxial
anisotropy, with equal anisotropy constant $K_{s},$ whose axes ${\bf e}_{i}$
are chosen to point outward and normal to the surface (\cite{Kodama}, \cite
{Dimitrov et al.}). More precisely, we define for each spin a unit gradient
vector on which the spin magnetic moment has to be projected. In the case of
a spherical particle these anisotropy axes are along the radius joining the
center of the particle to the considered surface site.
For an ellipsoidal particle the easy direction in the core is
taken along the major axis of the ellipsoid, which is also along the $z$
direction. The bulk anisotropy constant was estimated by many authors (see
e.g. \cite{Krupicka} and references therein) to be $K_{1}\simeq 4.7\times
10^{4}$ erg/cm$^{3}$. In our simulations we normalized this constant to the
number of sites and in units of K, $k_{c}\equiv (K_{c}/k_{B})\simeq
8.13\times 10^{-3}K,$ $k_{B}$ being the Boltzmann constant. On the other
hand, the surface anisotropy has not been determined experimentally in
magnetic oxides, but it has become clear, however, that the corresponding
contribution is very large as compared with the bulk one. In our
calculations we took $k_{s}\equiv (K_{s}/k_{B})\simeq 0.5$ K; the
corresponding surface anisotropy constant $K_{s}$ was estimated to be $%
\simeq 0.06$ erg/cm$^{2}$ (see \cite{Dormann et al.}, \cite{Aharoni2}). In 
\cite{Kodama} $k_{s}$ was taken in the range $1-4$ K, but for these higher
values of surface anisotropy the authors obtained high-field irreversibility
whereas the experimental results (also presented in \cite{Kodama}) show none
for the $\gamma $-Fe$_{2}$O$_{3}$ particles.

Finally, in the case of a very small ellipsoidal particle, as discussed at the
beginning of sect.2.A, the effect of the dipolar interactions ${\cal H}_{dip}
$ boils down to a mere shape anisotropy, which is absent in a spherical
particle. Therefore, the magnetostatic energy of an ellipsoid of semi-axes $%
\frac{X}{2},\frac{Y}{2},\frac{Z}{2},$ can be written as \cite{Akhiezer}, 
\cite{Aharoni} 
\begin{equation}
E_{mag}=\frac{1}{2V}\left( D_{x}\cdot M_{x}^{2}+D_{y}\cdot
M_{y}^{2}+D_{z}\cdot M_{z}^{2}\right)   \label{7}
\end{equation}
where $D_{\alpha },\alpha =x,y,z,$ are the demagnetizing factors, $M_{\alpha
}$ the components of the net magnetisation, and $V$ the
volume of the particle, which is equal to $N_{t}$ in our calculations. If
all semi-axes are different it is not possible to express the $D^{\prime }s$
in closed form. However, this is possible in the case of a prolate or oblate
spheroid. In the case of a prolate spheroid, as is very common in permanent-magnet 
materials, i.e. $Z>X=Y,$ the demagnetizing factors are given by \cite
{Scaife} 
\begin{eqnarray}
D_{z} &=&4\pi \frac{1-e^{2}}{2e^{3}}\left[ \log \left( \frac{1+e}{1-e}%
\right) -2e\right] ,  \label{8} \\
D_{x} &=&D_{y}=\frac{1}{2}(4\pi -D_{z}),  \nonumber
\end{eqnarray}
where $e=\sqrt{1-\left( X/Z\right) ^{2}},0<e<1,$ is the eccentricity of the
ellipsoid. In these calculations we assumed that the easy axes of the
magnetocrystalline and shape anisotropy are the same, though this is not the
case in general.

We have found that using the long-range dipolar interactions ${\cal H}_{dip}$ 
involving all possible pairs of atoms in the particle, or the macroscopic
magnetostatic energy $E_{mag}$ yields within numerical errors, the same
results, only that the former contribution is much more time consuming than
the latter.
\subsection{Results and discussion}
\label{Results}
\subsubsection{Thermal variation of the magnetisation}
\label{Thermal variation}
\begin{figure*}
\begin{center}
\resizebox{0.75\textwidth}{!}{%
  \includegraphics{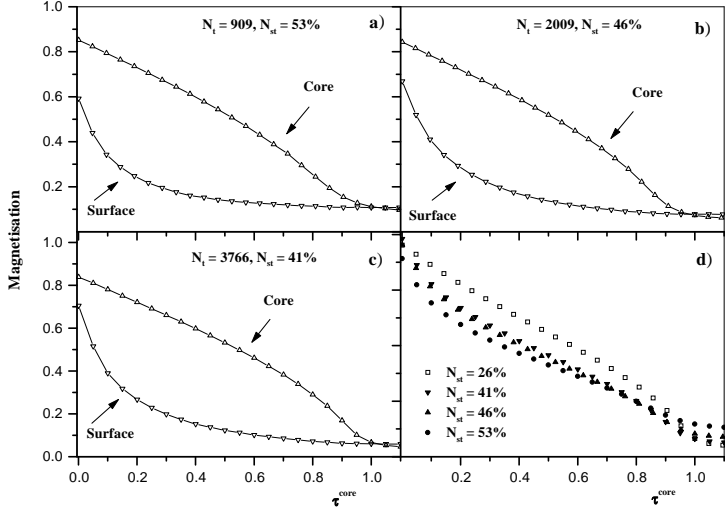}
}
\end{center}
\vspace*{-2.5cm}       
\caption{Thermal variation of the surface and core magnetisation (per site)
(Figs.3a-c) and mean magnetisation (Fig.3d) as
obtained from the Monte Carlo simulations of an ellipsoidal nanoparticle.
The anisotropy constants are given in the text; the exchange interactions on
the surface are taken to be $1/10$ times those in the core.}
\label{fig:3}       
\end{figure*}

In Figs.~\ref{fig:3}a-c, we plot the computed thermal variation of the core and surface
contributions to the magnetisation (per site) as a function of the reduced
temperature $\tau ^{core}\equiv T/T_{c}^{core}$, $T_{c}^{core}$ being the
highest core ``critical temperature'', for $N_{t}=909,2009,3766$ with 
$N_{st}=53\%,46\%,41\%$,
respectively. These values of $N_{st}$ have been determined by the fact that
the thickness of the surface shell is constant ($\sim 0.35$ nm), according
to M\"{o}ssbauer-effect analysis \cite{Tronc et al.}. They correspond to a diameter of 
circa $4,4.6$ and $6$ nm, respectively.

In Fig.~\ref{fig:3}d we plot the mean magnetisation defined as 
$M_{mean}\equiv (N_{s}M_{surface}+N_{c}M_{core})/N_t$ 
as a function of $\tau^{core},$ for the same values of $N_{st}$ as before.
The exchange coupling in the core (i.e. $J_{AA},J_{AB},J_{BB}$), generically
denoted by $J_{c},$ are taken as 10 times those on the surface, denoted by $%
J_{s}$. Since there is no experimental estimation of the exchange couplings
on the surface, the choice of such $J_{s}$ was guided by the fact that the
critical temperature of the bulk material is circa $906$ K, while the
hypothetical ``surface transition'' occurs in the temperature range of $30-75$
K (which is a factor of more than 10 smaller) according to M\"{o}ssbauer
spectroscopy and high-field magnetisation measurements. Nevertheless, we
have also considered the effect of other ratios $J_{c}/J_{s}$ (see below).
%
%
We see that the surface ``critical region'', corresponding to the would-be magnetic
phase transition on the surface is in a range of temperatures lower than the
critical temperature of the core. This is expected from the fact that since
the molecular field acting on a surface spin is lower than the one acting on
a core spin, as a consequence of the lower coordination number at the
surface and hence the change in crystal field thereon \cite{Wolfram et al.}, 
\cite{Eriksson et al.}, and also because of the fact that $J_{s}$ are taken
smaller than $J_{c}$. This is in agreement with the results of
M\"{o}ssbauer spectroscopy and the magnetisation measurements at high
fields, since the surface component has a ``critical temperature'' in the
range $30-75$ K while for the core component the critical temperature is
much higher ($\simeq 906$ K). Note that both transitions are smeared because 
of the finiteness of the system size (see discussion after Eq.(\ref{M})).
We also note that the surface magnetisation $M_{surface}$ decreases more
rapidly than the core contribution $M_{core}$ as the temperature increases,
and has a positive curvature while that of $M_{core}$ is negative, as is
most often the case. Moreover, it is seen that even the (normalized) core
magnetisation per site does not reach its saturation value of $1$ at very
low temperatures, and this can be explained by the fact that the magnetic
order in the core is disturbed by the relative disorder on the surface, or
in other words, the magnetic disorder starts at the surface and gradually
propagates into the core of the particle (see Fig.~\ref{fig:6} below). The
magnetic disorder on the surface is, of course, enhanced by the
single-site surface anisotropy which tends to orientate the spins normal
to the surface. 
In Fig.~\ref{fig:3}d we see
that the more important is the surface contribution the more enhanced and
rapid is the raising of the mean magnetisation at low temperatures, and this
behaviour bears some resemblance to Fig.~\ref{fig:1}c.

\begin{figure}
\resizebox{0.75\textwidth}{!}{%
\includegraphics{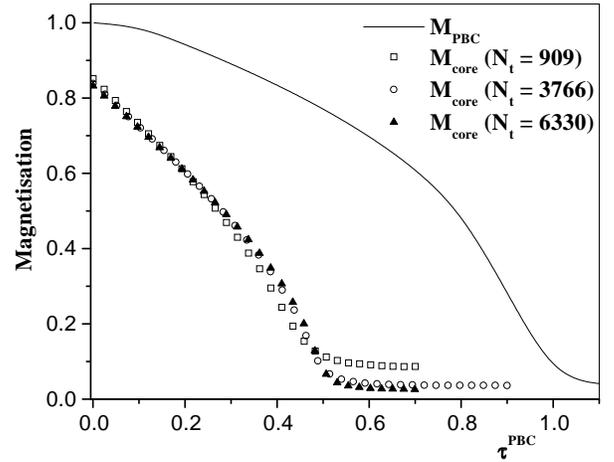}}
\vspace{-2.5cm}      
\caption{Thermal variation of the magnetisation of the PBC 
system with $N_{t} = 40^3$, and 
the core magnetisation for $N_{t}=909,3766,6330$ with $%
N_{st}=53\%,41\%,26\%,$ respectively, as functions of $\tau^{PBC}$ (see text). 
The exchange interactions on the surface are taken equal to $1/10$ times those 
in the core.}
\label{fig:4}       
\end{figure}

In Fig.~\ref{fig:4} we plot the core magnetisation of an ellipsoidal nanoparticle with 
$N_{t}=909,$ $3766,6330$ and the 
magnetisation of the isotropic system with the spinel structure and periodic boundary
conditions \footnote{%
This system is a perfectly ferrimagnetic material with periodic
spinel structure and without vacancies, though such material does not exist
in reality since all spinels present some degree of vacancy. This system will be
referred to in the sequel as the PBC system.}
as functions of the reduced temperature $\tau ^{PBC}\equiv T/T_{c}^{PBC},$
and $N_{st}=53\%,41\%,26\%$.
Comparing the different curves, it is seen that both the critical
temperature and the value of the magnetisation are dramatically
reduced in the core of the particle. The reduction of the critical
temperature is obviously
due to the finite-size and surface effects \cite{Binder & Krumbhaar}. There is a 
size-dependent
reduction of the critical temperature by up to 50\% for the smallest
particle. The same result has been found by Hendriksen et al.\cite
{Hendriksen et al.} for small clusters of various structures (bcc, fcc, and
disordered) using spin-wave theory. As to the magnetisation, the reduction
shows that the core of the particle does not exhibit the same magnetic
properties as the bulk material, and as discussed before, it is influenced by the
misaligned spins on the surface.
%
%
In Fig.~\ref{fig:4} we can also see that the higher $N_{t}$ the lower the magnetisation
in the critical region and the higher the temperature at which the
magnetisation approaches zero, and this is consistent with the fact that $M\propto 1/\sqrt{N_t}$
at high temperatures, as discussed earlier. However, the increase of the critical temperature 
with $N_{t}$
is not as clear-cut as it could be expected, and this can be understood by
noting that the disordered surface ($J_{s}=J_{c}/10,$ small coordination
numbers, and single-site anisotropy) strongly influences the magnetic order in the core through the
relatively strong exchange couplings between surface and core spins, which
are equal to those in the core.
\subsubsection{Effect of the ratio $J_{c}/J_{s}$}
\label{Effect of the ratio}
\begin{figure}
\resizebox{0.75\textwidth}{!}{%
\includegraphics{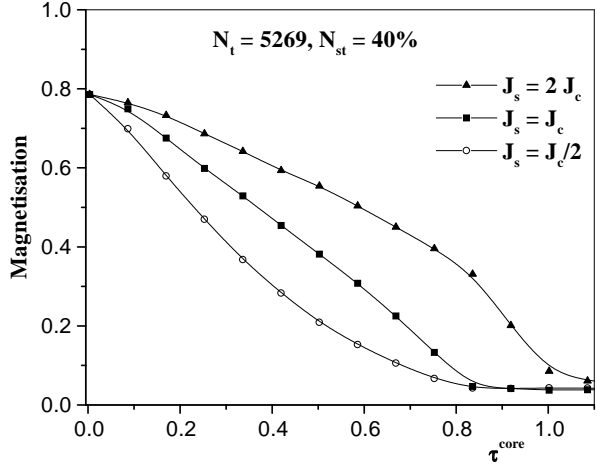}}
\vspace{-2.5cm}      
\caption{The computed thermal variation of the surface
contribution to the net magnetisation of an ellipsoidal nanoparticle as a
function of $\tau ^{core}$ for $J_{s}=J_{c}/2,J_{c},2J_{c}$ (see text).}
\label{fig:5}       
\end{figure}

In Fig.~\ref{fig:5} we plot the thermal variation of the surface 
magnetisation as a function of the reduced temperature $\tau ^{core}$ for
$J_{s}=J_{c}/2,J_{c}$ and $2J_{c},$ and $N_{t}=5269$, $%
N_{st}=40\%.$ Here the thickness of the surface shell is greater than $0.35$
nm, and is not taken from experiments, contrary to the values in Figs.~\ref{fig:3}. The
reason for taking such $N_{st}$ is merely to compare the results for
surfaces with different thickness. At any rate, our aim here is to maintain $%
N_{st}$ fixed to a given value and vary the ratio $J_{c}/J_{s}.$
%
%
We see that the surface critical region is shifted to higher
temperatures upon increasing $J_{s},$ and only when $J_{s}=2J_{c}$ that both
the core and surface magnetic ``phase transitions'' occur in the same
temperature range, i.e.\ $\tau ^{core}\simeq 1$. It is worth noting that the
weaker the exchange interactions on the surface the lower the
magnetisation of the latter.
This result remains the same upon lowering the surface width. In this case
the number of spins having smaller coordination numbers, and hence weaker
effective exchange energy, i.e. those spins on the outer shell of the
particle, is very small as compared with the rest of spins in the particle.
Moreover, we may think that the spins on the outer shell follow the (strong)
molecular field created by the other (inner) spins constituting a relatively
ferrimagnetically ordered core.
Setting $J_{c}=\lambda J_{s},$ we may determine the coefficient $\lambda $
at which the transition regions of the core and surface overlap. This may be
done using the simple, though time consuming, cumulant method introduced a
few years ago by Binder (see \cite{Binder & Krumbhaar} for a review).
\subsubsection{Profile of the magnetisation}
\label{Profile}
\begin{figure}
\resizebox{0.75\textwidth}{!}{%
\includegraphics{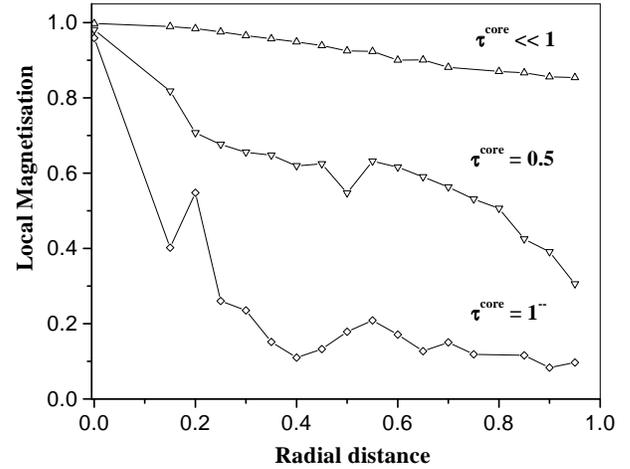}}
\vspace{-2.5cm}      
\caption{Spatial variation of the local magnetisation of a spherical nanoparticle of $3140$ 
spins, as a function of the radial distance, for $\tau^{core}\ll 1$, $\tau^{core} = 0.5$, 
and $\tau^{core} \simeq 1^-$.}
\label{fig:6}       
\end{figure}

We have determined the spatial variation of the local magnetisation of a
spherical particle of a fixed radius as we move from the center out onto
the surface, at
different temperatures. At nearly zero temperature ($\tau^{core} \ll 1$), the local
magnetisation decreases with increasing radial distance in the particle. It starts from the
saturation value of an iron atom (in fact, we computed the mean value of A
and B atoms, to take into account the possibility of starting at either
atom) at the center of the particle and decreases down to the value of a
surface atom. Note that what is plotted in Fig.~\ref{fig:6} is in fact the
(normalized) projection of the atomic magnetic moment along the easy axis
(z-axis for the core, and normal for the surface spins), which is
proportional to $\cos \theta $, and thus Fig.~\ref{fig:6} also shows the spatial
evolution of the orientation of the magnetic moment inside the particle as the radial 
distance is varied.
%
%
The decreasing of the local magnetisation confirms what was said before,
that is even at very low
temperature the surface is in a magnetic order which is different from that
in the core, and as discussed earlier, the misalignement of spins starts at the
surface and gradually propagates into the core. This could be related with
the gradual canting of spins confirmed by M\"{o}ssbauer spectroscopy \cite
{Haneda}, \cite{Coey}, \cite{Tronc et al.} even at 4.2 K. As the temperature increases 
($\tau^{core} = 0.5, \tau^{core} \simeq 1^-$), the local magnetisation exhibits 
a jump of temperature-dependent height, and continues to decrease.
Since the local magnetisation depends on the direction of the radius vector,
especially in an ellipsoidal particle, the curves in Fig.~\ref{fig:6} present, in
addition to the usual numerical inaccuracy (especially for $\tau^{core}
\simeq 1^-$), some fluctuations due to the
spatial non homogeneities inherent to the lacunous spinel structure of the $%
\gamma $-Fe$_{2}$O$_{3}$ particles ($\frac{1}{3}$ for each two B atoms,
randomly distributed).
\subsubsection{Specific heat}
\label{Cv}
\begin{figure}
\resizebox{0.75\textwidth}{!}{%
\includegraphics{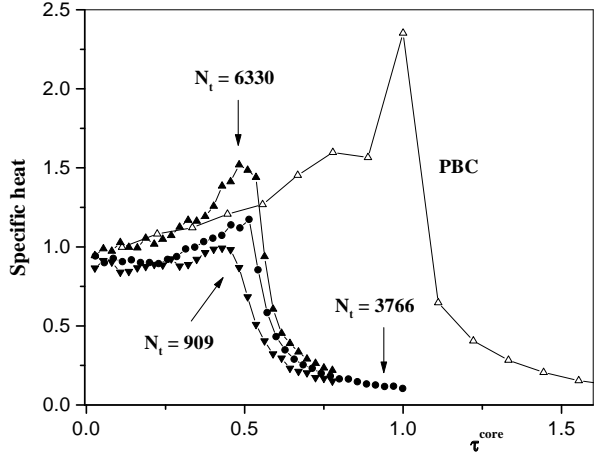}}
\vspace{-2.5cm}      
\caption{Thermal variation of the specific heat of the PBC system with $N_{t} = 40^3$, and
that of an ellipsoidal nanoparticle with $N_{t}=909,3766,6330$, and for the same energy
parameters as in Figs.~\ref{fig:3}).}
\label{fig:7}       
\end{figure}

In Fig.~\ref{fig:7} we plot the computed specific heat of the PBC system ($N_{t}=40^3$) 
and that of an ellipsoidal nanoparticle with $N_{t}=909,3766,6330$ (with the 
corresponding values of $N_{st}$ given earlier). 
There is a transition marked by a sharp peak for
the PBC system and a broad peak for the three particle sizes. This broadening is
an obvious illustration of the finite-size effects. Note also the shift by $%
50\%$ to lower temperatures of the critical region, as in the magnetisation
thermal behaviour in Fig.~\ref{fig:4}.
%
%
We are intending to prepare appropriate samples for the specific heat
measurements. To our knowledge, such measurements have never been performed on 
nanoparticle assemblies.
\section{Conclusion}
\label{Conclusion}%
We have presented a microscopic model for magnetic nano-scale particles including
exchange and dipolar interactions, and bulk and surface anisotropy, and
investigated the thermal and spatial behaviours of the core and surface
magnetisation of a nanoparticle of different sizes,
and hence of different surface contributions. We have found that the
finiteness of the system size and free boundaries lead to a non uniform
magnetisation profile decreasing toward the surface in the particle. Moreover,
it turns out that surface anisotropy also leads to non saturation of the magnetisation
at low temperatures.

In order to compare our results obtained for a single nanoparticle with
our experiments on dilute assemblies of nanoparticles \cite{Ezzir}, and in particular to
compare with the curve of magnetisation vs. temperature given in Fig.~\ref{fig:1}b, we
have to include the Zeeman energy term in our Hamiltonian and run our
program for different particle sizes with randomly distributed easy axes.
Unfortunately, this will require several days of CPU time, since a
temperature sweep takes about 30 hours (without dipolar interactions) on a
two-processor Alpha Work Station. 
It also remains to study the interplay between finite-size and surface effects in a 
nanoparticle of round shape (sphere or ellipsoid), as is the case here,
since for a box-shaped particle one can treat separately finite-size and
surface effects by considering both cases of free boundaries where these
two effects are mixed and that of periodic boundary conditions where only
the former are present.
In the present case, it can be seen from Fig.~\ref{fig:4} that the magnetisation of the PBC system 
where only finite-size effects are present is larger than the bulk magnetisation, while 
the magnetisation of a (anisotropic) nanoparticle is lower. This implies, as was analytically shown 
in \cite{HKDG} for a box-shaped particle with simple-cubic structure, that finite-size effects 
yield a positive contribution to the magnetisation while surface effects render a larger and 
negative contribution, resulting in a net magnetisation that is lower than that of the bulk system. 
In \cite{HK} it was also shown that the difference between the finite-size and surface contributions 
is enhanced by surface anisotropy, which leads to non saturation of the magnetisation at low 
temperatures (see Figs.~\ref{fig:3}-\ref{fig:5}).
Nevertheless, already from the present single-particle preliminary study we
can infer some conclusions about the effect of the magnetically disordered surface 
on the global magnetic properties of nanoparticles. The surface contribution to the
magnetisation presents a rather different behaviour from the core
contribution. Furthermore, even at very low temperatures, the local
magnetisation decreases with the distance from the center, showing that the
magnetic state on the surface is definitely different from that in the core. 
We have also shown that there is a drastic reduction of both the
critical temperature and the value of the magnetisation in the core of the
particle. The reduction of the critical temperature is obviously due to the finite-size
and surface effects. As to the magnetisation, the reduction shows that the
magnetic properties of the particle core are different from those of bulk
material, because of the strong influence of the misalignement of spins on 
the surface, driven by symmetry breaking of the cristal field and surface
anisotropy, which propagates from the boundary into the core.
Finally, we note that our results on the thermal and spatial behaviour of the
magnetisation are in qualitative agreement with those obtained by Binder et al. 
\cite{Wildpaner} for spherical particles with simple-cubic crystalline
structure.
\bigskip

{\bf Acknowledgments}

HK\ thanks L.\ Reynaud for his considerable help in these simulations and
D.\ Garanin for his valuable remarks and suggestions. The present numerical
calculations have been performed on the Alpha Work Station of the Genome
Laboratory of the University of Versailles to which we are greatly
endebted.

\end{document}